\documentclass[11pt,a4paper]{article}
\usepackage[hyperref]{acl2020}
\usepackage{graphicx,graphics}
\usepackage{subcaption}
\usepackage{multirow}
\usepackage{booktabs}
\usepackage{times}
\usepackage{amsfonts,amsmath}
\usepackage{latexsym}

\usepackage{comment}

\usepackage{soul}

\usepackage{microtype}

\aclfinalcopy 
\def\aclpaperid{***} 

\title{Improved Speech Representations\\with Multi-Target Autoregressive Predictive Coding}

\author{Yu-An Chung, James Glass\\
  Computer Science and Artificial Intelligence Laboratory\\
  Massachusetts Institute of Technology\\
  Cambridge, MA 02139, USA\\
  \texttt{\{andyyuan,glass\}@mit.edu}\\}

\date{}

\begin{document}
\maketitle
\begin{abstract} 
Training objectives based on predictive coding have recently been shown to be very effective at learning meaningful representations from unlabeled speech.
One example is Autoregressive Predictive Coding~\citep{chung2019unsupervised}, which trains an autoregressive RNN to generate an unseen future frame given a context such as recent past frames.
The basic hypothesis of these approaches is that hidden states that can accurately predict future frames are a useful representation for many downstream tasks.
In this paper we extend this hypothesis and aim to enrich the information encoded in the hidden states by training the model to make more accurate future predictions.
We propose an auxiliary objective that serves as a regularization to improve generalization of the future frame prediction task.
Experimental results on phonetic classification, speech recognition, and speech translation not only support the hypothesis, but also demonstrate the effectiveness of our approach in learning representations that contain richer phonetic content.
\end{abstract}

\section{Introduction}


Unsupervised speech representation learning, which aims to learn a function that transforms surface features, such as audio waveforms or spectrograms, to higher-level representations using only unlabeled speech, has received great attention recently~\citep{baevski2019effectiveness,baevski2020vq,liu2020mockingjay,song2019speech,jiang2019improving,schneider2019wav2vec,chorowski2019unsupervised,pascual2019learning,oord2018representation,kamper2019truly,chen2018phonetic,chung2018speech2vec,chung2018unsupervised,milde2018unspeech,chung2016audio,hsu2017unsupervised}.
A large portion of these approaches leverage self-supervised training, where the learning target is generated from the input itself, and thus can train a model in a {\it supervised} manner.

\citet{chung2019unsupervised} propose a method called Autoregressive Predictive Coding (APC), which trains an RNN to predict a future frame that is $n$ steps ahead of the current position given a context such as the past frames.
The training target can be easily generated by right-shifting the input by $n$ steps.
Their intuition is that the model is required to produce a good summarization of the past and encode such knowledge in the hidden states so as to accomplish the objective.
After training, the RNN hidden states are taken as the learned representations, and are shown to contain speech information such as phonetic and speaker content that are useful in a variety of speech tasks~\citep{chung2020generative}.

Following their intuition, in this work we aim to improve the generalization of the future frame prediction task by adding an auxiliary objective that serves as a regularization.
We empirically demonstrate the effectiveness of our approach in making more accurate future predictions, and confirm such improvement leads to a representation that contains richer phonetic content.

The rest of the paper is organized as follows.
We start with a brief review of APC in Section~\ref{sec:apc-review}.
We then introduce our approach in Section~\ref{sec:proposed-approach}.
Experiments and analysis are presented in Section~\ref{sec:experiments}, followed by our conclusions in Section~\ref{sec:conclusions}.

\begin{figure*}[ht]
  \centering
  \includegraphics[width=0.65\textwidth]{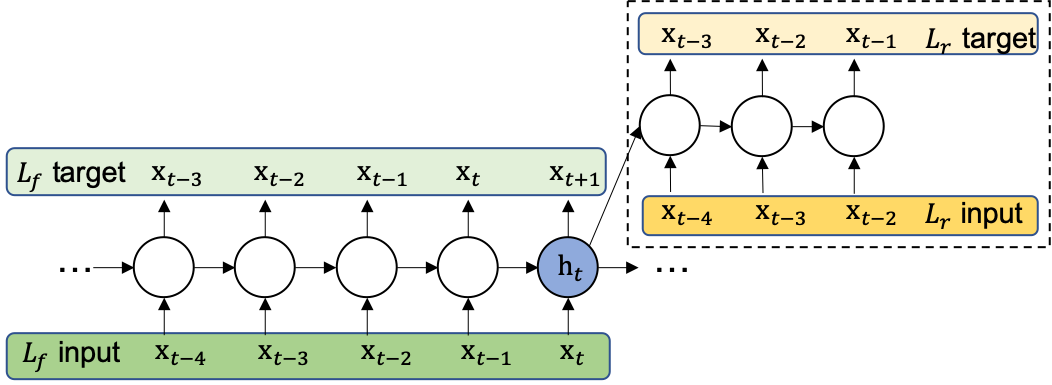}
  \caption{Overview of our method. $L_{f}$ is the original APC objective that aims to predict $x_{t + n}$ given a context $(x_{1}, x_{2}, \dots, x_{t})$ with an autoregressive RNN. Our method first samples an anchor position, assuming it is time step $t$. Next, we build an auxiliary loss $L_{r}$ that computes $L_{f}$ of a past sequence $(x_{t - s}, x_{t - s + 1}, \dots, x_{t - s + \ell - 1})$ (see Section~\ref{sec:remembering} for definitions of $s$ and $\ell$), using an auxiliary RNN (dotted line area). In this example, $(n, s, \ell) = (1, 4, 3)$. In practice, we can sample multiple anchor positions, and averaging over all of them gives us the final $L_{r}$.}
  \label{fig:multitask-apc}
\end{figure*}

\section{Autoregressive Predictive Coding}
\label{sec:apc-review}
Given a context of a speech signal represented as a sequence of acoustic feature vectors $(x_{1}, x_{2}, \dots, x_{t})$, the objective of Autoregressive Predictive Coding (APC) is to use the context to infer a future frame $x_{t + n}$ that is $n$ steps ahead of $x_{t}$.
Let $\mathbf{x} = (x_{1}, x_{2}, \dots, x_{N})$ denote a full utterance, where $N$ is the sequence length, APC incorporates an RNN to process each frame $x_{t}$ sequentially and update its hidden state $h_{t}$ accordingly.
For $t = 1, \dots, N - n$, the RNN produces an output $y_{t} = \mathbf{W}\cdot h_{t}$, where $\mathbf{W}$ is an affinity matrix that maps $h_{t}$ back to the dimensionality of $x_{t}$.
The model is trained by minimizing the frame-wise L1 loss between the predicted sequence $(y_{1}, y_{2}, \dots, y_{N - n})$ and the target sequence $(x_{1 + n}, x_{2 + n}, \dots, x_{N})$:
\begin{equation}
  L_{f}(\mathbf{x}) = \sum_{t = 1}^{N - n}|x_{t + n} - y_{t}|.
  \label{eq:predict-future}
\end{equation}

When $n = 1$, one can view APC as an {\it acoustic} version of neural LM (NLM)~\citep{mikolov2010recurrent} by thinking of each acoustic frame as a token embedding, as they both use a recurrent encoder and aim to predict information about the future.
A major difference between NLM and APC is that NLM infers tokens from a closed set, while APC predicts frames of real values.

Once an APC model is trained, given an utterance $(x_{1}, x_{2}, \dots, x_{N})$, we follow~\citet{chung2019unsupervised} and take the output of the last RNN layer $(h_{1}, h_{2}, \dots, h_{N})$ as its extracted features.

\section{Proposed Methodology}
\label{sec:proposed-approach}

Our goal is to make APC's prediction of $x_{t + n}$ given $h_{t}$ more accurate.
In Section~\ref{sec:experiments} we will show this leads to a representation that contains richer phonetic content.

\subsection{Remembering more from the past}
\label{sec:remembering}
An overview of our method is depicted in Figure~\ref{fig:multitask-apc}.
We propose an auxiliary loss $L_{r}$ to improve the generalization of the main objective $L_{f}$ (Equation~\ref{eq:predict-future}).

The idea of $L_{r}$ is to refresh the current hidden state $h_{t}$ with the knowledge learned in the past.
At time step $t$, we first sample a past sequence $\mathbf{p}_{t} = (x_{t - s}, x_{t - s + 1}, \dots, x_{t - s + \ell - 1})$, where $s$ is how far the start of this sequence is from $t$ and $\ell$ is the length of $\mathbf{p}_{t}$.
We then employ an auxiliary RNN, denoted as RNN$_{\mathrm{aux}}$, to perform predictive coding defined in Equation~\ref{eq:predict-future} conditioning on $h_{t}$.
Specifically, we initialize the hidden state of RNN$_{\mathrm{aux}}$ with $h_{t}$, and optimize it along with the corresponding $\mathbf{W}_{\mathrm{aux}}$ using $L_{f}(\mathbf{p}_{t})$, which equals to $\sum_{t' = t - s}^{t - s + \ell - 1}|x_{t' + n} - y_{t'}|$.
Such a process reminds $h_{t}$ of what has been learned in $h_{t - s}, h_{t - s + 1}, \dots, h_{t - s + \ell - 1}$.

For a training utterance $\mathbf{x} = (x_{1}, x_{2}, \dots, x_{N})$, we select each frame with probability $P$ as an anchor position.
Assume we end up with $M$ anchor positions: $a_{1}, a_{2}, \dots, a_{M}$.
Each $a_{m}$ defines a sequence $\mathbf{p}_{a_{m}} = (x_{a_{m} - s}, x_{a_{m} - s + 1}, \dots, x_{a_{m} - s + \ell - 1})$ before $x_{a_{m}}$, which we use to compute $L_{f}(\mathbf{p}_{a_{m}})$.
Averaging over all anchor positions gives the final auxiliary loss $L_{r}$:
\begin{equation}
  L_{r}(\mathbf{x}) = \frac{1}{M}\sum_{m = 1}^{M} L_{f}(\mathbf{p}_{a_{m}}).
  \label{eq:reconstruct-past}
\end{equation}
The final APC objective combines Equations~\ref{eq:predict-future} and~\ref{eq:reconstruct-past} with a balancing coefficient $\lambda$:
\begin{equation}
  L_{m}(\mathbf{x}) = L_{f}(\mathbf{x}) + \lambda L_{r}(\mathbf{x}).
  \label{eq:multitask-loss}
\end{equation}
We re-sample the anchor positions for each $\mathbf{x}$ during each training iteration, while they all share the same RNN$_{\mathrm{aux}}$ and $\mathbf{W}_{\mathrm{aux}}$.

\section{Experiments}
\label{sec:experiments}
We demonstrate the effectiveness of $L_{r}$ in helping optimize $L_{f}$, and investigate how the improvement is reflected in the learned representations.

\begin{figure*}[t]
  \centering
  \begin{subfigure}[t]{0.5\textwidth}
    \centering
    \includegraphics[height=1.8in]{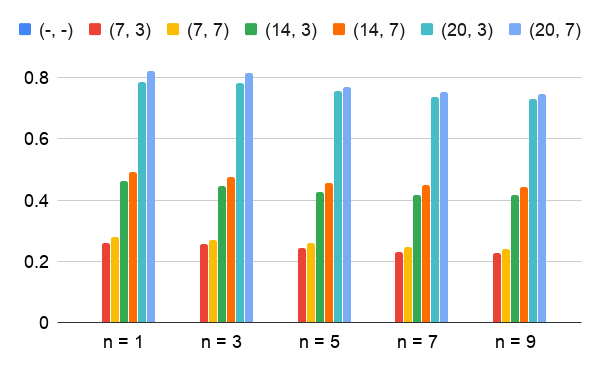}
    \caption{$L_{r}$ (auxiliary objective, Equation~\ref{eq:reconstruct-past})}
    \label{fig:lr-plot}
  \end{subfigure}%
  \begin{subfigure}[t]{0.5\textwidth}
    \centering
    \includegraphics[height=1.8in]{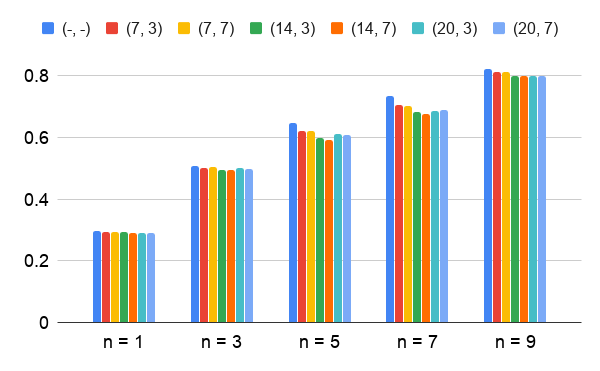}
    \caption{$L_{f}$ (main objective, Equation~\ref{eq:predict-future})}
    \label{fig:lf-plot}
  \end{subfigure}
  \caption{Validation loss of $L_{r}$ (left) and $L_{f}$ (right) on LibriSpeech {\small \texttt{dev-clean}} when training APC using different $(n, s, \ell)$ combinations. Each bar of the same color represents one $(s, \ell)$ combination. We use $(-, -)$ to denote an APC optimized only with $L_{f}$.
Bars are grouped by their $n$'s with different $(s, \ell)$ combinations within each group.}
\end{figure*}

\subsection{Setup}
We follow~\citet{chung2019unsupervised} and use the audio portion of the LibriSpeech~\citep{panayotov2015librispeech} {\small \texttt{train-clean-360}} subset, which contains 360 hours of read speech produced by 921 speakers, for training APC.
The input features are 80-dimensional log Mel spectrograms, i.e., $x_{t}\in \mathbb{R}^{80}$.
Both RNN and RNN$_\mathrm{aux}$ are a 3-layer, 512-dim unidirectional GRU~\citep{cho2014properties} network with residual connections between two consecutive layers~\citep{wu2016google}.
Therefore, $\mathbf{W}, \mathbf{W}_\mathrm{aux}\in \mathbb{R}^{512\times 80}$.
$\lambda$ is set to 0.1 and the sampling probability $P$ is set to 0.15, that is, each frame has a 15\% of chance to be selected as an anchor position.
$P$ and $\lambda$ are selected based on the validation loss of $L_f$ on a small data split.
All models are trained for 100 epochs using Adam~\citep{kingma2015adam} with a batch size of 32 and a learning rate of $10^{-3}$.

\subsection{Effect of $L_{r}$}
We first validate whether augmenting $L_{r}$ improves $L_{f}$.
As a recap, $n$ is the number of time steps ahead of the current position $t$ in $L_{f}$, and $s$ and $\ell$ denote the start and length, respectively, of a past sequence before $t$ to build $L_{r}$.
We consider $(n, s, \ell) \in \{1, 3, 5, 7, 9\} \times \{7, 14, 20\} \times \{3, 7\}$.
Note that each phone has an average duration of about 7 frames.

Figures~\ref{fig:lr-plot} and~\ref{fig:lf-plot} present $L_{r}$ (before multiplying $\lambda$) and $L_{f}$ of the considered APC variants on the LibriSpeech {\small \texttt{dev-clean}} subset, respectively.
Each bar of the same color represents one $(s, \ell)$ combination.
We use $(-, -)$ to denote an APC optimized only with $L_{f}$.
Bars are grouped by their $n$'s with different $(s, \ell)$ combinations within each group.

We start with analyzing Figure~\ref{fig:lr-plot}.
Note that $L_{r}$ does not exist for $(-, -)$ and is set to 0 in the figure.
We see that under the same $n$, the performance of $L_{r}$ is mainly decided by how far ($s$) the past sequence is from the current position rather than the length ($\ell$) to generate: when we keep $\ell$ fixed and increase $s$ from 7 (red), 14 (green), to 20 (blue), we observe the loss surges as well.

From Figure~\ref{fig:lf-plot}, we have the following findings.
\paragraph{For a small $n$, the improvement in $L_{f}$ brought by $L_{r}$ is minor.}
By comparing $(-, -)$ with other bars, we see that when $n\leq 3$, which is smaller than half of the average phone duration (7 frames),
adding $L_{r}$ does not lower $L_{f}$ by much.
We speculate that when $n\leq 3$, $x_{t + n}$ to be inferred is usually within the same phone as $x_{t}$, making the task not challenging enough to force the model to leverage more past information.

\paragraph{$L_{r}$ becomes useful when $n$ gets larger.}
We see that when $n$ is close to or exceeds the average phone duration ($n\geq 5$), an evident reduction in $L_{f}$ after adding $L_{r}$ is observed, which validates the effectiveness of $L_{r}$ in assisting with the optimization of $L_{f}$.
When $n = 9$, the improvement is not as large as when $n = 5$ or $7$.
One possible explanation is that $x_{t + 9}$ has become almost independent from the previous context $h_{t}$ and hence is less predictable.

By observing the validation loss, we have shown that $L_{r}$ indeed helps generalize $L_{f}$.

\begin{table*}[htbp]
  \footnotesize  
  \centering
    \begin{tabular}{lccccccc}
      \toprule
      \multirow{2}{*}{Feature}  &  \multicolumn{7}{c}{Time shift}\\
      \cmidrule(lr){2-8}
                         &  -15   &  -10   &   -5   &    0   &   +5   &  +10   &  +15\\
      \midrule
      \midrule
      log Mel            &  83.3  &  80.3  &  67.6  &  49.9  &  65.5  &  77.9  &  82.7\\
      \midrule
      \multicolumn{8}{c}{APC trained with $L_{f}$ (Equation~\ref{eq:predict-future})}\\
      \midrule
      $n = 1$            &  56.1  &  45.8  &  36.1  &  33.7  &  56.5  &  73.7  &  81.6\\
      $n = 3$            &  50.8  &  41.8  &  34.8  &  33.4  &  56.0  &  73.5  &  81.1\\
      $n = 5$            &  48.7  &  38.2  &  32.5  &  31.9  &  54.8  &  73.0  &  80.5\\
      $n = 7$            &  44.6  &  38.6  &  32.9  &  32.1  &  56.3  &  73.8  &  80.4\\
      $n = 9$            &  51.0  &  41.8  &  35.7  &  36.9  &  58.4  &  74.6  &  81.0\\
      \midrule
      \multicolumn{8}{c}{APC trained with $L_{m}$ (Equation~\ref{eq:multitask-loss})}\\
      \midrule
      $n = 1$            &  50.6  &  42.2  &  35.1  &  33.1  &  54.4  &  73.4  &  81.4\\
      $n = 3$            &  46.4  &  38.0  &  34.1  &  32.4  &  54.1  &  71.4  &  80.5\\
      $n = 5$            &  41.8  &  35.1  &  29.8  &  28.1  &  49.6  &  64.6  &  76.8\\
      $n = 7$            &  39.8  &  33.8  &  28.7  &  27.8  &  46.8  &  60.6  &  74.4\\
      $n = 9$            &  42.3  &  35.3  &  30.3  &  29.7  &  50.0  &  63.3  &  76.6\\
      \bottomrule
    \end{tabular}
  \caption{Phonetic classification results using different types of features as input to a linear logistic regression classifier. The classifier aims to correctly classify each frame into one of the 48 phone categories. Frame error rates ($\downarrow$) are reported. Given a time shift $w\in \{0, \pm 5, \pm 10, \pm 15\}$, the classifier is asked to predict the phone identity of $x_{t + w}$ given $x_{t}$.}
  \label{tab:phone-classification}
\end{table*}

\subsection{Learned representation analysis}
Next, we want to examine whether an improvement in $L_{f}$ leads to a representation that encodes more useful information.
Speech signals encompass a rich set of acoustic and linguistic properties.
Here we will only focus on analyzing the phonetic content contained in a representation, and leave other properties such as speaker for future work.

We use phonetic classification on TIMIT~\citep{garofolo1993darpa} as the probing task to analyze the learned representations.
The corpus contains 3696, 400, and 192 utterances in the train, validation, and test sets, respectively.
For each $n\in \{1, 3, 5, 7, 9\}$, we pick the $(s, \ell)$ combination that has the lowest validation loss.
As described in Section~\ref{sec:apc-review}, we take the output of the last RNN layer as the extracted features, and provide them to a linear logistic regression classifier that aims to correctly classify each frame into one of the 48 phone categories.
During evaluation, we follow the protocol~\citep{lee1989speaker} and collapse the prediction to 39 categories.
We report frame error rate (FER) on the test set, which indicates how much phonetic content is contained in the representations.
We also conduct experiments for the task of predicting $x_{t - w}$ and $x_{t + w}$ given $x_{t}$ for $w\in \{5, 10, 15\}$.
This examines how contextualized $h_{t}$ is, that is, how much information about the past and future is encoded in the current feature $h_{t}$.
We simply shift the labels in the dataset by $\{\pm 5, \pm 10, \pm 15\}$ and retrain the classifier.
We keep the pre-trained APC RNN fixed for all runs.
Results are shown in Table~\ref{tab:phone-classification}.

We emphasize that our hyperparameters are chosen based on $L_f$ and are never selected based on their performance on any downstream task, including phonetic classification, speech recognition, and speech translation to be presented next.
Tuning hyperparameters towards a downstream task defeats the purpose of unsupervised learning.

\paragraph{Phonetic classification}
We first study the standard phonetic classification results, shown in the column where time shift is 0.
We see that APC features, regardless of the objective ($L_{f}$ or $L_{m}$), achieve lower FER than log Mel features, showing that the phonetic information contained in the surface features has been transformed into a more accessible form (defined as how linearly separable they are).
Additionally, we see that APC features learned by $L_{m}$ outperform those learned by $L_{f}$ across all $n$.
For $n\geq 5$ where there is a noticeable improvement in future prediction after adding $L_{r}$ as shown in Figure~\ref{fig:lf-plot}, their improvement in phonetic classification is also larger than when $n\leq 3$.
Such an outcome suggests that APC models that are better at predicting the future do learn representations that contain richer phonetic content.
It is also interesting that when using $L_{f}$, the best result occurs at $n = 5$ (31.9); while with $L_{m}$, it is when $n = 7$ that achieves the lowest FER (27.8).

\paragraph{Predicting the past or future}
We see that it is harder to predict the nearby phone identities from a log Mel frame, and the FER gets higher further away from the center frame.
An APC feature $h_{t}$ contains more information about its past than its future.
The result matches our intuition as the RNN generates $h_{t}$ conditioning on $h_{i}$ for $i < t$ and thus their information are naturally encoded in $h_{t}$.
Furthermore, we observe a consistent improvement in both directions by changing $L_{f}$ to $L_{m}$ across all $n$ and time shifts.
This confirms the use of $L_{r}$, which requires the current hidden state $h_{t}$ to recall what has been learned in previous hidden states, so more information about the past is encoded in $h_{t}$.
The improvement also suggests that an RNN can forget the past information when training only with $L_{f}$, and adding $L_{r}$ alleviates such problem.

\subsection{Speech recognition and translation}
The above phonetic classification experiments are meant for analyzing the phonetic properties of a representation.
Finally, we apply the representations learned by $L_{m}$ to automatic speech recognition (ASR) and speech translation (ST) and show their superiority over those learned by $L_{f}$.

We follow the exact setup in~\citet{chung2020generative}.
For ASR, we use the Wall Street Journal corpus~\citep{paul1992design}, use {\small \texttt{si284}} for training, and report the word error rate (WER) on {\small \texttt{dev93}}.
For ST, we use the LibriSpeech En-Fr corpus~\citep{kocabiyikoglu2018augmenting}, which aims to translate an English speech to a French text, and report the BLEU score~\citep{papineni2002bleu}.
For both tasks, the downstream model is an end-to-end, sequence-to-sequence RNN with attention~\citep{chorowski2015attention}.
We compare different input features to the same model.
Results, shown in Table~\ref{tab:asr-ast}, demonstrate that the improvement in predictive coding brought by $L_{r}$ not only provides representations that contain richer phonetic content, but are also useful in real-world speech applications.%
\footnote{According to~\citet{chung2020generative}, when using a Transformer architecture~\citep{vaswani2017attention,liu2018generating} as the autoregressive model, representations learned with $L_{f}$ can achieve a WER of 13.7 on ASR and a BLEU score of 14.3 on ST.}

\begin{table}[htbp]
  \footnotesize
  \centering
  \begin{tabular}{lcc}
    \toprule
    Feature         &  ASR (WER $\downarrow$)  &  ST  (BLEU $\uparrow$)\\
    \midrule
    \midrule
    log Mel         &           18.3           &  12.9\\
    APC w/ $L_{f}$  &           15.2           &  13.8\\
    APC w/ $L_{m}$  &           14.2           &  14.5\\
    \bottomrule
  \end{tabular}
  \caption{Automatic speech recognition (ASR) and speech translation (ST) results using different types of features as input to a seq2seq with attention model. Word error rates (WER, $\downarrow$) and BLEU scores ($\uparrow$) are reported for the two tasks, respectively.}
  \label{tab:asr-ast}
\end{table}

\section{Conclusions}
\label{sec:conclusions}
We improve the generalization of Autoregressive Predictive Coding by multi-target training of future prediction $L_{f}$ and past memory reconstruction $L_{r}$, where the latter serves as a regularization.
Through phonetic classification, we find the representations learned with our approach contain richer phonetic content than the original representations, and achieve better performance on speech recognition and speech translation.



\bibliography{acl2020}
\bibliographystyle{acl_natbib}

\end{document}